\documentclass[twocolumn,showpacs,prl,superscriptaddress,floatfix]{revtex4}

\usepackage{amsmath} 
\usepackage{amsfonts} 
\usepackage{amssymb}

\usepackage{graphicx} 
\usepackage{epsfig} 
\usepackage{subfigure}

\usepackage{tabularx} 
\usepackage{wasysym} 
\usepackage{dcolumn}
\usepackage{bm} 
\usepackage{psfrag} 
\usepackage{times}

\usepackage[svgnames]{xcolor} 
\usepackage[pdftex]{hyperref}
\hypersetup{	colorlinks=true,
		linkcolor=blue,
		citecolor=ForestGreen,
		urlcolor=DarkOrchid
           }

\newcommand{\harvard}{\affiliation{Department of Chemistry and Chemical Biology, Harvard University, 12 Oxford Street, 02138 Cambridge, Massachusetts, USA}} 
\newcommand{\sgt}{\affiliation{Stinger Ghaffarian
Technologies Inc., 7701 Greenbelt Rd., Suite 400, Greenbelt, MD 20770}} 
\newcommand{\nasa}{\affiliation{NASA Ames Research Center Quantum Artificial Intelligence Laboratory (QuAIL), Mail Stop 269-1, 94035 Moffett Field CA}} 
\newcommand{\tamu}{\affiliation{Department of Physics and Astronomy, Texas A\&M University, College Station, Texas 77843-4242, USA}}

\newcommand{\sfi}{\affiliation{Santa Fe Institute, 1399 Hyde Park Road, Santa Fe, New Mexico 87501 USA}}

\begin{document}

\title{Exponentially Biased Ground-State Sampling of Quantum Annealing
Machines\\ with Transverse-Field Driving Hamiltonians}

\author{Salvatore Mandr{\`a}} \email{smandra@fas.harvard.edu} \harvard \nasa \sgt

\author{Zheng Zhu} \email{zzwtgts@tamu.edu} \tamu

\author{Helmut G.~Katzgraber} \email{hgk@tamu.edu} \tamu \sfi

\date{\today}

\begin{abstract}

We study the performance of the D-Wave 2X quantum annealing machine on
systems with well-controlled ground-state degeneracy. While obtaining
the ground state of a spin-glass benchmark instance represents a
difficult task, the gold standard for any optimization algorithm or
machine is to sample all solutions that minimize the Hamiltonian with
more or less equal probability. Our results show that while n\"aive
transverse-field quantum annealing on the D-Wave 2X device can find the
ground-state energy of the problems, it is not well suited in
identifying all degenerate ground-state configurations associated to a
particular instance. Even worse, some states are exponentially
suppressed, in agreement with previous studies on toy model problems
[New J.~Phys.~{\bf 11}, 073021 (2009)]. These results suggest that more
complex driving Hamiltonians are needed in future quantum annealing
machines to ensure a fair sampling of the ground-state manifold.

\end{abstract}


\maketitle

Optimization is ubiquitous across disciplines. Finding optimization
approaches that quickly and reliably estimate the ground state of a
complex optimization problem is of great importance. While many
algorithmic approaches from computer science have had a great impact in
physics problems, similarly, physics-inspired optimization techniques
have revolutionized optimization in fields as broad as engineering,
biology, chemistry, and computer science, to name a few. One
physically inspired optimization technique that has found widespread
application is simulated annealing \cite{kirkpatrick:83}. Similar to
thermal annealing invented towards the end of the neolithic era
\cite{comment:otzi}, the heuristic is straightforward to implement.
Initially, the system is prepared at a high temperature and it is left
to thermalize. The temperature is sequentially reduced and, during the
process, the system is enforced, if possible, to be in thermal
equilibrium at any given temperature. At the end of the annealing
(namely, when a specific target temperature is reached), the lowest
energy configuration recorded during the process is returned.
\begin{figure}[t!]
\includegraphics[width=\columnwidth]{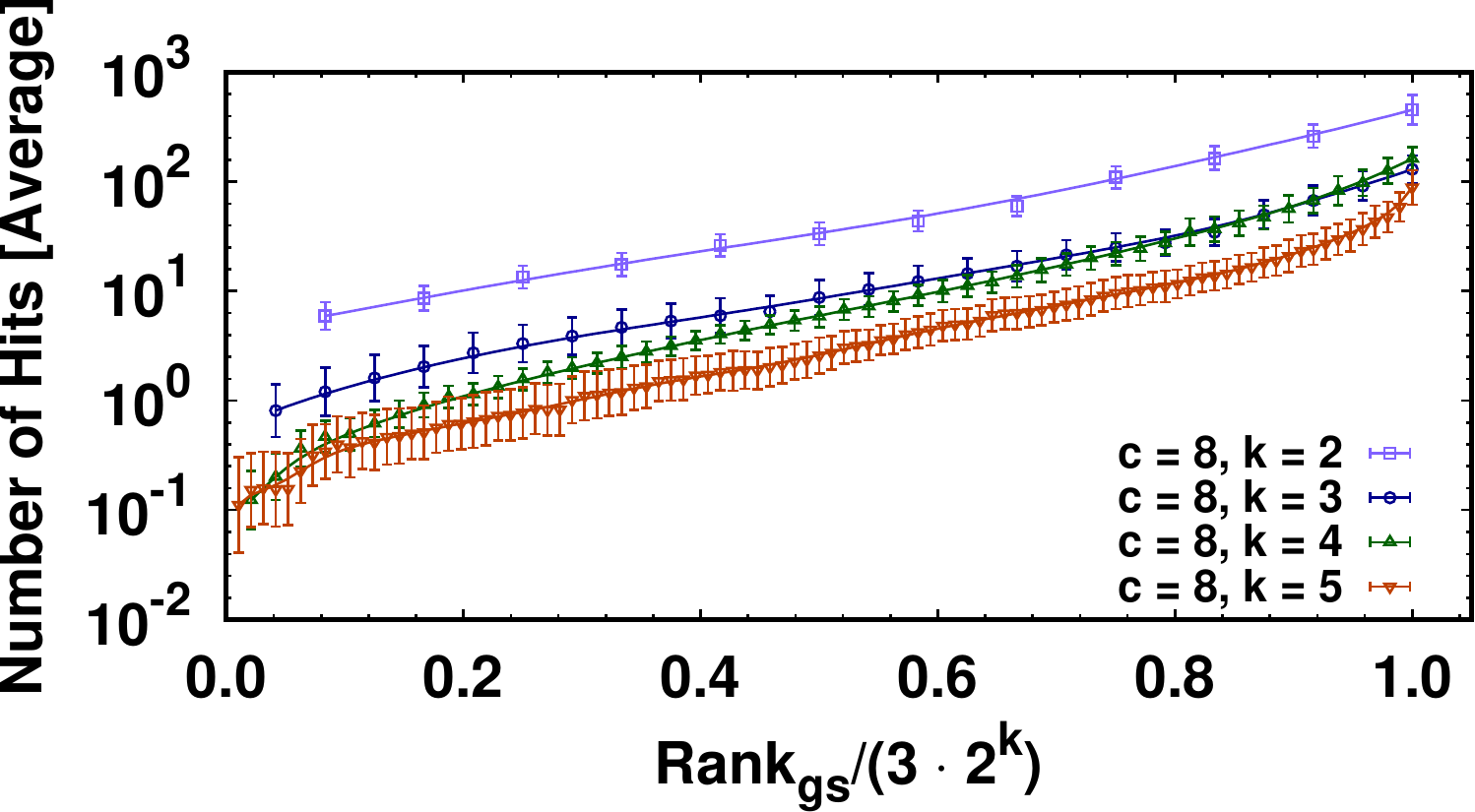}\\
\includegraphics[width=\columnwidth]{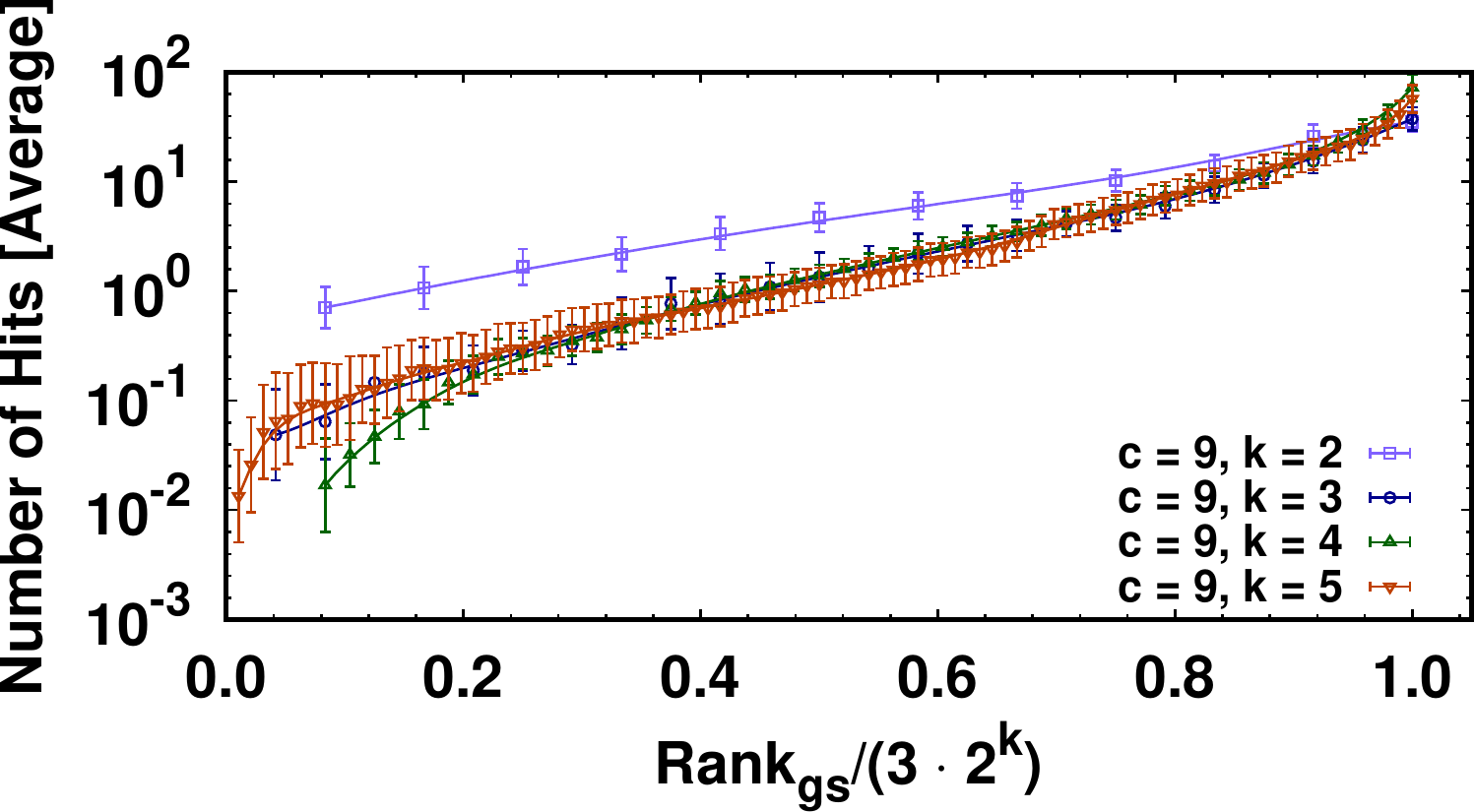}
\caption{\label{fig:ngs}
Histograms of the number of times a particular ground-state
configuration is found using the DW2X quantum annealer ($20\,\mu s$
annealing time) sorted by rank. Data for Chimera lattice instances with
$N = 8 c^2$ sites and $N_{\rm GS} = 3\times2^k$ ground states are
shown. The horizontal axis is normalized by $N_{\rm GS}$ for easier
display of the data. In all cases studied, certain ground states are
exponentially suppressed (note the vertical logarithmic axis).}
\end{figure}
The procedure is repeated with different initial conditions to ensure that
the obtained state is, actually, the lowest-energy state. Most
importantly, it has been shown rigorously that simulated annealing can
indeed obtain the ground state of a system for sufficiently long
annealing \cite{geman:84}; however, this is not practical.
Nevertheless, it often fails to find the global minimum when the energy
landscape of the problem Hamiltonian has many metastable states, such as
is the case of, e.g., spin glasses
\cite{binder:86,mezard:87,young:98,stein:13}. More recently, the quantum
counterpart of simulated annealing (usually called ``quantum
annealing'') was suggested
\cite{finnila:94,kadowaki:98,brooke:99,farhi:01,santoro:02,das:05,santoro:06,das:08,morita:08}.
In this case, quantum fluctuations are typically induced by a transverse
field (instead of thermal fluctuations) to drive transitions from state
to state. The advantage of quantum annealing is that the induced quantum
fluctuations, in principle, could aid in the search for the optimum by
allowing the system to tunnel across thin energy barriers. To date, it
remains controversial if it is able to outperform simulated annealing or
other classical optimization methods.

Interest in quantum annealing has been considerably boosted by the
introduction of the D-Wave quantum annealers \cite{comment:d-wave}.
These devices experimentally implement finite-temperature quantum
annealing with a transverse field on a system of Boolean variables
coupled together on a topology known as the Chimera graph
\cite{bunyk:14}. Advantages in the use of the method beyond
specially crafted problems for Chimera's architecture
\cite{venturelli:15a,denchev:16,mandra:16b} remain to be found, the
D-Wave 2X (DW2X) machine can be considered a huge technological feat
with radically new technology.  Interestingly, while many aspects of the
DW2X have been scrutinized in detail, no detailed tests on its ``fair
sampling'' \cite{moreno:03,matsuda:09} abilities---namely, the ability to
sample all states of a degenerate problem with (hopefully) equal
probability---have been performed.  Studies on toy problems and simple
Hamiltonians suggest that transverse-field driven quantum annealing does
not uniformly sample all the possible ground states resulting in some
configuration being exponentially suppressed \cite{matsuda:09}. Studies
on different generations of the D-Wave quantum annealer
\cite{boixo:13a,albash:15a,king:16} already suggested that the sampling
might be biased, but no systematic study has been performed to date.
This can be seen as a noticeable shortcoming of the optimization
technique.

\begin{figure}[t!]
\includegraphics[width=\columnwidth]{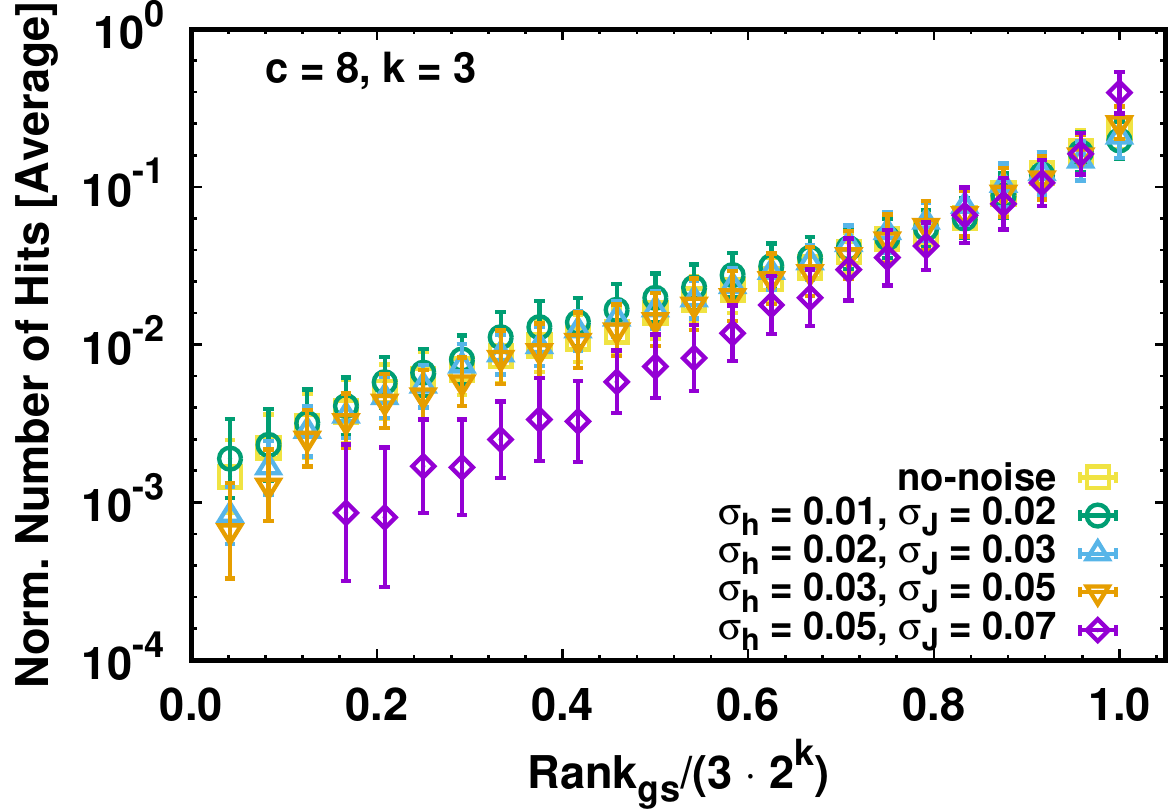}\\
\caption{\label{fig:noise}
Histogram of the number of times a particular ground-state configuration
is found using the DW2X quantum annealer ($20\,\mu s$ annealing time)
sorted by rank, after adding extra Gaussian noise to the couplers with
variance $\sigma_{J}^2$ and biases with variance
$\sigma_{h}^2$.  Data for Chimera lattice instances with $N = 8
c^2$ sites and $N_{\rm GS} = 3\times2^k$ ground states are shown. The
horizontal axis is normalized by $N_{\rm GS}$ for easier display of the
data. In all cases studied, noise minimally affects the exponential bias
of the sampling (note the vertical logarithmic axis).}
\end{figure}

So why is the exponential suppression of certain states, i.e., the lack
of fair sampling, such a problem? First, because good optimization
techniques should deliver all possible configurations that minimize the
problem Hamiltonian (provided enough repetitions and using different
initial conditions) in addition to being fast and reliable. This
encompasses a far more stringent quality test for any optimizer. Second,
and most importantly, there are many important applications for which a
fair sampling of states is fundamental.  In physical applications, a
fair sampling of states is imperative when estimating the ground-state
entropy of a degenerate system.  Similarly in computer science, for many
combinatorial problems, if one can sample uniformly from the set of
solutions, then one can use these different solutions to obtain a
highly accurate count of the total number of solutions \cite{jerrum:86},
which is important for propositional model counting (\#SAT)
\cite{gomes:08} and the knapsack solution counting problem (\#Knapsack)
\cite{gopalan:11}.  Finally, in multiple industrial applications having
many different solutions to a problem is highly desirable. For example,
many uncorrelated solutions are needed to construct probabilistic
membership filters using SAT formulas \cite{weaver:14,douglass:15}. As
such, a quantum annealing machine with a transverse-field driving
Hamiltonian might not be the best approach to solve these problems. On
one hand, one can hope that the inherent noise found in the analog DW2X
might help alleviate these biases of transverse-field quantum annealing.
On the other hand, this problem could be alleviated with more complex
driving Hamiltonians \cite{matsuda:09}. Unfortunately, such machines are
only being constructed at the moment.

In this Letter we demonstrate experimentally that, for spin-glass problems
with a small (known) number of ground-state configurations, the DW2X is
heavily biased towards some configurations, while other minimizing
configurations are exponentially suppressed. Despite applying multiple
gauges, performing many runs, or increasing the annealing time, the
machine is unable to sample the states fairly; i.e., it is not well
suited for a wide variety of optimization applications.

{\em Description of the benchmark instances.---} We perform the
experiments on the D-Wave Systems, Inc., DW2X quantum annealing
machine \cite{comment:d-wave}. We use all operable qubits on the machine
and encode spin-glass problems on the couplers
\cite{binder:86,nishimori:01,stein:13} of the underlying Chimera
topology of the system \cite{bunyk:14}. The Hamiltonian of the problem
is ${\mathcal H} =  -\sum_{\{i,j\} \in {\mathcal V}}J_{ij}S^z_i S^z_j$.
The $N$ Ising variables $S^z_i \in \{\pm 1\}$ are defined on the
vertices ${\mathcal V}$ of the Chimera lattice of size $N = 8 c^2$
(with $c \in \{8,9,10,11\}$) and do not couple to any local fields
(biases). The sum is over all edges ${\mathcal E}$ connecting vertices
$\{i,j\} \in {\mathcal V}$. Note that some couplers and/or qubits are
always inoperable. The aforementioned system sizes are for the complete
lattices without taking into account any defects.

To perform a controlled study of the effects of ground-state degeneracy,
we carefully choose the couplings from a Sidon set \cite{katzgraber:15}
with $J_{ij} \in \{\pm 5, \pm 6, \pm 7\}$.  Furthermore, after randomly
placing the couplings, we recursively traverse the lattice and shuffle
the interactions randomly so that no spins have a zero local field.
This prevents any additional degeneracy due to a larger number of free
spins \cite{katzgraber:15,zhu:16}. Because of our choice of disorder, we
find that the randomly generated instances have typically a ground-state
degeneracy of $N_{\rm GS} = 3\times 2^k$ ($k \in {\mathbb N}$). Some
instances have values of $N_{\rm GS}$ that do not fall into the sequence
$N_{\rm GS} = \{6, 12, 24, 48, 96, \ldots\}$ because of the
imperfections in the Chimera graph. We choose not to use such instances
for the experiments to perform a systematic study. Note that for small
subsections of the Chimera graph, i.e., for small $c$, the number of
ground states is typically smaller than for the largest possible lattice
with $N = 1152$ ($c=12$) sites \cite{comment:dead}. Therefore, the
available values of $k$ are smaller.

{\em Experimental details.---} The number of ground states for each
problem is determined classically using the isoenergetic cluster
algorithm (ICA) \cite{zhu:15b,zhu:16,zhu:16x,zhu:16y,houdayer:01}, which
is known to sample the ground-state manifold fairly, especially for
small numbers of ground states (here, small means $N_{\rm GS} \lesssim
10^3$).  ICA combines parallel tempering Monte Carlo simulations with
isoenergetic cluster moves (simulation parameters are shown in Table
\ref{tab:params_1}). To ensure that the lowest energy state has been
found, we independently simulate four system replicas with the same
couplings. More precisely, we check that the lowest energy found by each
replica (considering only the lowest temperatures) in $N_{\rm sw}/2$
updates, with $N_{\rm sw}$ the total number of updates, agree.  Hence, we
claim that the ground state has been found and we begin to record the
ground-state configurations, and the corresponding frequencies, for the
remaining $N_{\rm{\rm sw}}/2$ updates. There is no guarantee that any
solution obtained by this heuristic method is the true optimum, or that
we have found all configurations that minimize the Hamiltonian. However,
we ensure each configuration achieves a minimum number of $50$ hits in
order to increase our confidence that all accessible ground states have
been found. Moreover, we also check that the lowest energy is in
agreement with the Hamze--de Freitas--Selby heuristic
\cite{hamze:04,selby:14}.

\begin{figure}[t!]
\includegraphics[width=\columnwidth]{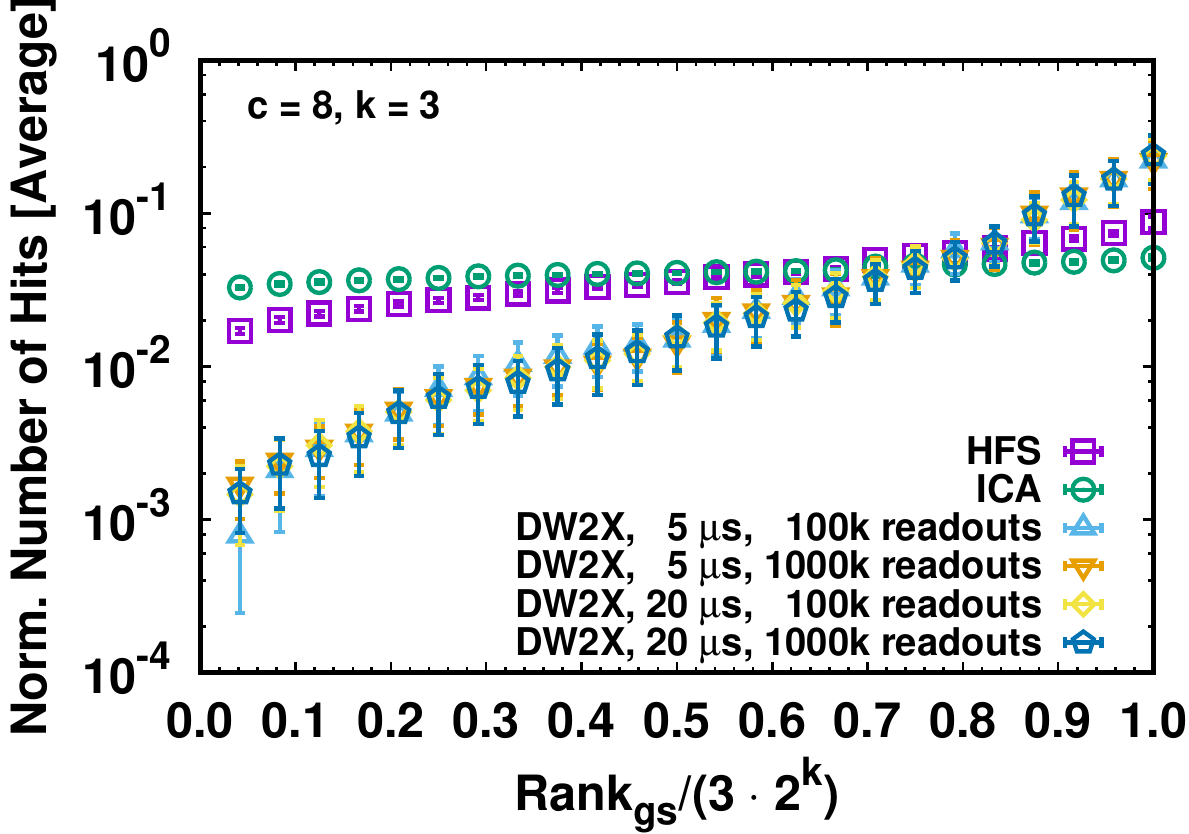}
\caption{\label{fig:comparison}
Histogram of the number of times a particular ground-state configuration
is found either using classical heuristics [using the isoenergetic
cluster algorithm (ICA) \cite{zhu:15b} and the Hamze--de Freitas--Selby
algorithm (HFS) \cite{hamze:04,selby:14}], as well as the DW2X quantum
annealer. Example data for Chimera lattice instances with $N = 8 
c^2$ sites and $N_{\rm GS} = 3\times2^k$ ground-states.  While the
classical algorithms sample the ground-state configurations fairly, the
DW2X device has a bias of more than 2 orders of magnitude.}
\end{figure}

\begin{table}
\caption{Simulation parameters for the isoenergetic cluster algorithm
(ICA): for each system size $N$, we compute $N_{\rm sa}$ disorder
instances. $N_{\rm sw} = 2^b$ is the total number of Monte Carlo sweeps
for each of the $4 N_T$ replicas for a single instance, $T_{\rm min}$
[$T_{\rm max}$] is the lowest [highest] temperature simulated, and $N_T$
is the number of temperatures used in the parallel tempering method. For
the lowest $N_{\rm ICA}$ temperatures isoenergetic cluster moves are
applied.  \label{tab:params_1}}
\begin{tabular*}{\columnwidth}{@{\extracolsep{\fill}} l l l l l l r }
\hline
\hline
$N$ & $N_{\rm sa}$ & $b$ & $T_{\rm min}$ & $T_{\rm max}$ & $N_{T}$ &$N_{\rm ICA}$  \\
\hline
$512$  &  $4164$ & $19$ & $0.06$ & $3.05$ & $33$ &$18$\\
$648$  &  $6970$ & $19$ & $0.06$ & $3.05$ & $33$ &$18$\\
$800$  & $11\,199$ & $19$ & $0.06$ & $3.05$ & $33$ &$18$\\
$968$  & $16\,739$ & $19$ & $0.06$ & $3.05$ & $33$ &$18$\\
\hline
\end{tabular*}
\end{table}

Quantum annealing experiments have been performed on the DW2X using a
fixed annealing time of $20~\mu s$. For each instance, we used $100$
distinct gauges and $1000$ readouts per gauge, for a total number of
$10^5$ readouts per instance.

\begin{figure}[t!]
\includegraphics[width=\columnwidth]{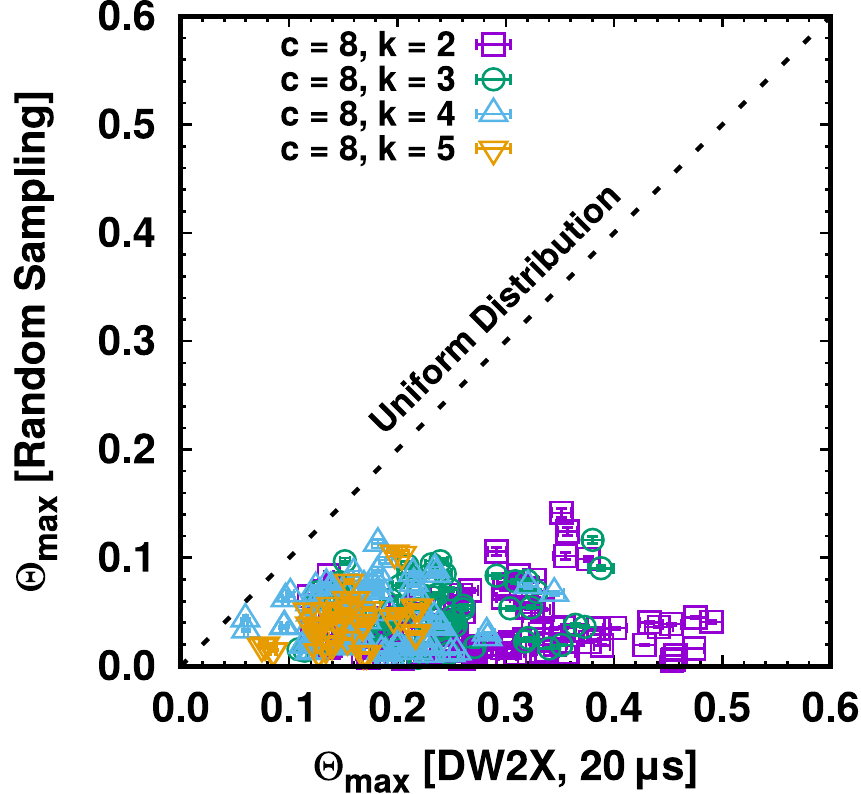}
\caption{\label{fig:scatter_theta}
Comparison of $\Theta_\text{max}$ (maximum absolute difference of the
empiric cumulative distribution with respect to the expected
distribution) for the distribution of ground states found by the DW2X
against a uniform random distribution. Each point corresponds to a
specific instance and the error bars are computed by bootstrapping the
data. The closer the points are to the diagonal, the more the
ground states for that specific instance have been uniformly sampled. As
one can see, DW2X does not uniformly sample the ground states.  The
example data are for Chimera lattices with $N = 8 c^2$ sites and
$N_\text{GS} = 3 \times 2^k$ ground states. For the analysis, we
considered only those instances for which DW2X found at least $50$ solutions
(independently of the ground-state configuration).
}
\end{figure}

{\em Results.---} Figure~\ref{fig:ngs} summarizes our results. Each
panel shows a histogram with the number of times a particular
ground-state configuration is found by the DW2X.  The horizontal axis
represents the index of a given ground-state configuration, normalized
by $3\times 2^k$ for a better readability. For each instance, indexes of
ground states are ordered so that the ground states with the largest
probability have the largest index.  Each panel represents different
experiments at a fixed system size $N = 8 c^2$, while each line
considers only experiments with a fixed number of ground-state
configurations $N_{\rm GS} = 3\times2^k$. Error bars are computed by
averaging each bin over a given number of samples.  In all cases
studied, some ground-state configurations are exponentially suppressed
(note the vertical logarithmic axis). We obtain similar results by
increasing the annealing time to $200\,\mu s$.  It is important to
stress that the exponential bias is minimally affected by introducing
additional artificial noise to the target Hamiltonian, as shown in
Fig.~\ref{fig:noise}.  Both random biases and coupler noise are drawn
from a Gaussian distribution with variances $\sigma_{h}^2$ and
$\sigma_{J}^2$, respectively.  In addition, we compare the sampling
of the DW2X to the two most efficient classical heuristics in
Fig.~\ref{fig:comparison}. While the bias is minimal for the classical
approaches (due to Poissonian fluctuations \cite{moreno:03}), a bias of
approximately 2 orders of magnitude persists for the DW2X device.

Finally, to better appreciate the exponential suppression of some ground states
of the DW2X, we introduce the observable $\Theta_\text{max}$ defined as
the maximum absolute difference of the empiric cumulative distribution
$\tilde F(x)$ with respect to the cumulative of a uniform distribution
$F(x)$, namely, $\Theta_\text{max} = \max_{x} |\tilde F(x) - F(x)|$, with
$x$ the ground-state index.  The test (which is similar in the purpose
of the Kolmogorov-Smirnov test) is useful to understand how close an
empiric distribution is to the expected distribution. More precisely,
the smaller $\Theta_\text{max}$ is, the more similar the distributions
are.  In Fig.~\ref{fig:scatter_theta}, we show the comparison of
$\Theta_\text{max}$ for the distribution of ground states found by the
DW2X against random numbers uniformly chosen in the set $\{1,\ 2,\
\ldots,\ N_\text{GS}\}$.  In general, the number of ground states that
DW2X can find widely varies from instance to instance.  Therefore, to
perform a fairer analysis, we extract an amount of random numbers which
is equal to the number of solutions (regardless of the ground-state
configurations) that the DW2X has found for the given instance. Each
point in the plot corresponds to a specific instance and the error bars
are computed by bootstrapping the data after the randomization of the
ground-state indices. The diagonal line represents the best value that
$\Theta_\text{max}$ can assume: the closer the points are to the
diagonal, the more uniformly the ground states for that specific
instance have been sampled.  For the analysis, we considered only those
instances for which DW2X has found at least $50$ solutions
(independently of the ground-state configuration). As one can see, the
results show that all the considered instances are far from the optimal
diagonal, which confirms that the DW2X using a transverse-field
driving Hamiltonian does not sample uniformly.  In addition, results
from instances with fixed $c$ and different $k$ suggest that the DW2X 
slightly improves its sampling by increasing the total number
of ground states. An intuitive understanding of how degeneracy of
ground states changes sampling can be obtained by considering level
crossings between ground states and low-energy excited states: instances
with less degeneracy tend to be harder and more likely to have level
crossings \cite{boixo:14,katzgraber:15}; therefore, a longer annealing
time is required to reach a stationary distribution of ground states.
Instances with larger degeneracy, however, have a slightly better fair
sampling for the same amount of annealing time.

\begin{figure}[t!]
\includegraphics[width=\columnwidth]{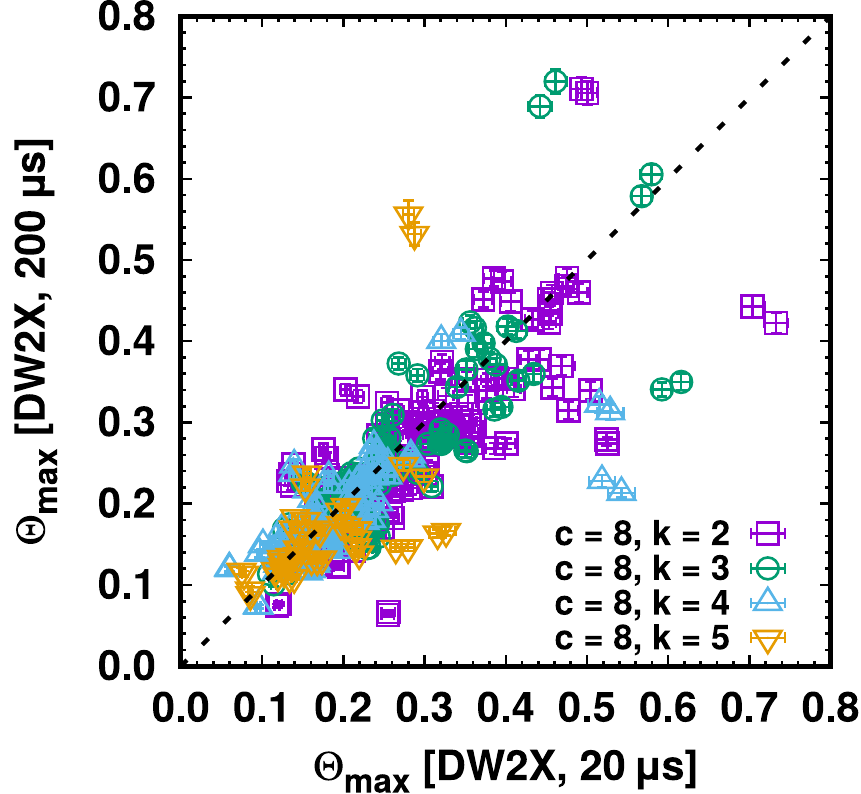}
\caption{\label{fig:20-200}
Comparison of $\Theta_\text{max}$ (defined as the maximum absolute
difference of the empiric cumulative distribution with respect to the
expected one) for the distribution of ground states found by the DW2X
for two different annealing times, $20\,\mu s$ and $200\,\mu s$. Each
point in the plot corresponds to a specific instance and the error bars
are computed by bootstrapping the data.  Only instances for which DW2X
has found a solution for both annealing times are represented. The data
are for Chimera lattice instances with $N = 8 c^2$ sites and
$N_\text{GS} = 3 \times 2^k$ ground states.  Even though there is a
slight improvement by increasing the total annealing time,
$\Theta_\text{max}$ remains quite large, thus showing that sampling
remains strongly biased even when the annealing time is increased
tenfold.}
\end{figure}

The abysmal fair-sampling performance of transverse-field quantum
annealing on the DW2X suggests that the machine is not well suited for
applications where many uncorrelated optimal states are needed.
Surprisingly, neither the intrinsic thermal fluctuations nor the
application of multiple gauges seem to affect these results
\cite{comment:apo}. Attempting to run the machine for longer annealing
times (see Fig.~\ref{fig:20-200}) has a negligible effect on the poor
sampling of the machine. This is in agreement with simulations on simple
toy models \cite{matsuda:09}. There, simulations showed that the use of
more complex driving Hamiltonians might alleviate this problem.
Finally, changing the energy scale of the Hamiltonian in the device, as
well as adding additional artificial noise, does not affect the poor
sampling (see Fig.~\ref{fig:noise}). As such, and in agreement with the aforementioned
analytical results, the transverse-field driver is likely the source of
the bias.  Unfortunately, at the moment neither quantum annealing
machines with more complex driving Hamiltonians nor quantum Monte Carlo
simulations to emulate these are readily available. However, the
aforementioned results strongly argue for more complex annealing
architectures in future devices.

{\em Summary.---} We have demonstrated experimentally that the D-Wave 2X
quantum annealer is unable to fairly sample states of degenerate random
spin-glass problems. In fact, some states are exponentially suppressed
compared to others. This means that transverse-field quantum annealing
might not be well suited for applications where many uncorrelated
solutions are needed. This could also explain the poor performance of
the implementation of probabilistic membership filters on the D-Wave
device \cite{douglass:15}. Our results are in agreement with previous
theoretical and numerical studies \cite{matsuda:09} on toy models and
suggest that the ever-growing quantum annealing community should put
more emphasis on mitigating this problem by, e.g., using more complex
driving Hamiltonians \cite{matsuda:09} or developing hybrid
architectures that encourage thermal fluctuations \cite{dickson:13}. We
do emphasize, however, that degenerate embedded problems might be
affected differently by this problem. For example, different embeddings
might influence the sampling differently, a problem that should be
studied in the future.

We would like to thank A.~Aspuru-Guzik, F.~Hamze, A.~King, A.~J.~Ochoa
and A.~Perdomo-Ortiz for fruitful discussions.  We also thank
E.~G.~Rieffel and D.~Venturelli for carefully reading the manuscript.
H.G.K.~acknowledges support from the NSF (Grant No.~DMR-1151387) and
would like to thank Zaya for inspiration to initiate this project.
S.M.~was supported by NASA (Sponsor Award No.~NNX14AF62G). We thank the
Texas Advanced Computing Center (TACC) at The University of Texas at
Austin for providing HPC resources (Stampede Cluster) and Texas A\&M
University for access to their Ada and Lonestar clusters. The research
of H.G.K.~and Z.Z.~is based upon work supported in part by the Office of
the Director of National Intelligence (ODNI), Intelligence Advanced
Research Projects Activity (IARPA), via MIT Lincoln Laboratory Air Force
Contract No.~FA8721-05-C-0002.  The views and conclusions contained
herein are those of the authors and should not be interpreted as
necessarily representing the official policies or endorsements, either
expressed or implied, of ODNI, IARPA, or the U.S.~Government.

The U.S.~Government is authorized to reproduce and distribute reprints
for Governmental purpose notwithstanding any copyright annotation
thereon. All authors contributed equally to the project.

\bibliography{refs,comments}

\begin{thebibliography}{44}
\bibitem[{com({\natexlab{a}})}]{comment:otzi}
\bibinfo{note}{For example, one of the objects found with {\"O}tzi the iceman
  was a copper axe.}
\bibitem[{com({\natexlab{c}})}]{comment:dead}
\bibinfo{note}{$N = 1152$ is the theoretical maximal size of the chip. However,
  trapped fluxes and manufacturing imperfections reduce the problem size to
  approximately $100$ sites.}
\bibitem[{com({\natexlab{d}})}]{comment:apo}
\bibinfo{note}{At this time it is unclear of the intrinsic noise of the D-Wave
  device either helps with the fair sampling or makes it worse. However, given
  the results presented in Ref.~\cite{matsuda:09}, we expect that noise should
  not be the sole source of the lack of fair sampling.}
\end{thebibliography}

\end{document}